\documentclass[twocolumn,showpacs,amsmath,amssymb,pra]{revtex4}
\usepackage{bbm}

\usepackage{graphicx}
\usepackage{dcolumn}
\usepackage{bm}
\usepackage{color}
\usepackage{mathrsfs}
\usepackage{epstopdf}
\usepackage{ulem}
\usepackage{float}
\usepackage{tensor}

\newcommand{\be}{\begin{equation}}
\newcommand{\ee}{\end{equation}}
\newcommand{\bey}{\begin{eqnarray}}
\newcommand{\eey}{\end{eqnarray}}
\newcommand{\bw}{\begin{widetext}}
\newcommand{\ew}{\end{widetext}}

\newcommand{\ba}{\begin{array}}
\newcommand{\ea}{\end{array}}
\newcommand{\bi}{\begin{itemize}}
\newcommand{\ei}{\end{itemize}}
\newcommand{\bem}{\begin{enumerate}}
\newcommand{\eem}{\end{enumerate}}

\newcommand{\avg}[1]{\left\langle #1 \right\rangle}
\newcommand{\ket}[1]{\left| #1 \right\rangle}
\newcommand{\set}[1]{\left\{ #1 \right\}}
\newcommand{\evol}[3]{\left\langle #1 \left| #2 \right| #3 \right\rangle}

\begin{document}

\title {Quantum work distributions associated with the dynamical Casimir effect}

%
%

\author{Zhaoyu Fei}
\affiliation{School of Physics, Peking University, Beijing 100871, China}
\author{Jing-Ning Zhang}\email[Email:]{jnzhang13@mail.tsinghua.edu.cn}
\affiliation{Center for Quantum Information, Institute for Interdisciplinary Information Sciences, Tsinghua University, Beijing 100084, China}
\author{Rui Pan}
\affiliation{School of Physics, Peking University, Beijing 100871, China}
\author{Tian Qiu}
\affiliation{School of Physics, Peking University, Beijing 100871, China}
\author{H. T. Quan}\email[Email:]{htquan@pku.edu.cn}
\affiliation{School of Physics, Peking University, Beijing 100871, China}
\affiliation{Collaborative Innovation Center of Quantum Matter, Beijing 100871, China}

 \date{\today}

 \begin{abstract}

We study the joint probability distribution function of the work and the change of photon number of the nonequilibrium process of driving the electromagnetic (EM) field in a three-dimensional cavity with an oscillating boundary. The system is initially prepared in a grand canonical equilibrium state and we obtain the analytical expressions of the characteristic functions of work distributions in the single-resonance and multiple-resonance conditions. Our study demonstrates the validity of the fluctuation theorems of the grand canonical ensemble in nonequilibrium processes with particle creation and annihilation. In addition, our work illustrates that in the high temperature limit, the work done on the quantized EM field approaches its classical counterpart; while in the low temperature limit, similar to Casimir effect, it differs significantly from its classical counterpart.

 \end{abstract}

 \maketitle


 \section{Introduction}

 Fluctuation theorems have attracted considerable attentions in the field of nonequilibrium thermodynamics in the last two decades. One of the most important results is the Jarzynski equality~\cite{no1997}
 \be
 \langle e^{-\beta w}\rangle=e^{-\beta \Delta F}.
 \ee
 It connects the equilibrium free energy difference $\Delta F$ with the fluctuating nonequilibrium work $w$ done on a system initially prepared in a canonical equilibrium state at the inverse temperature $\beta=1/k_BT$. This equality is first derived in the classical regime and later generalized to the quantum regime~\cite{aq2000,ja2000,wo2007}. The validity has been tested experimentally in various systems~ \cite{eq2002,ve2005,ex2015}.

 For the grand canonical equilibrium initial state, besides the trajectory work $w$, the particle number along every ``trajectory" is also a fluctuating quantity and the Jarzynski equality takes a similar form~\cite{st2007,th2009,wo2012}
 \be
 \langle e^{-\beta( w-\mu \Delta N)}\rangle=e^{-\beta \Delta \Phi},
 \ee
 where $\mu$ is the chemical potential of the initial state, $\Delta N$ and $\Delta \Phi$ are the change of the particle numbers and the difference of the grand potentials respectively before and after the force protocol. This equality has been discussed in chemical reaction networks~\cite{st2007}, exchange fluctuation systems~\cite{qu2011,cl2004,th2009,sy2008} and isolated systems without particle number change~\cite{wo2011,no2011,wo2012}. However fluctuation theorems of the grand canonical ensemble in isolated systems in nonequilibrium processes with particles creation and annihilation have not been studied so far (but see Ref.~\cite{ja2018}). The reason may be that usually the mass gap of massive particles is so large that the energy input during the driving of the boundary is too small to create massive particles.

 Nevertheless, photons are massless particles, which makes it possible to create photons with relatively low energy input. One example is the dynamical Casimir effect (DCE). The DCE is a quantum effect which describes the generation of photons due to 
 the EM field in the presence of time-dependent boundaries. The study was initiated by Moore~\cite{qu1970} and followed by many researchers~ \cite{fulling1976,ef1994,cole1995,ge1996,tr1998,no2001,re2001,qu2002,he2005,th2006,ph2006,dodonov2010,dalvit2011,nation2012}. The first experimental verification was carried out in 2011~\cite{johansson2010,ob2011}. The researchers used a modified SQUID to mimic a mirror moving at the required relativistic velocity and observed the DCE in a superconducting circuit. The main concern in the studies of DCE is about the average number of photon creation. The studies of the DCE inspire us to investigate the distributions of work and photon number change in the nonequilibrium processes with the photon creation and annihilation. It is worth mentioning that classical ideal gas insider a cylinder~\cite{lua2005} has been a prototype model for the study of thermodynamics. EM field in a cavity is an analogue~\cite{quan2009,ja2009} of gas insider a cylinder, but with more complicated dynamics, because it incorporates effects of quantum mechanics, quantum statistics, and special relativity. Nevertheless, for certain specific protocols, we are able to obtain the analytical results of the joint distribution of work and the change of photon number. The analytical results enable us to study the quantum-classical correspondence of the work done on the EM field in the high temperature limit and the quantum nature (Casimir effect) of the work in the low temperature limit.

 In this paper, we focus on the DCE in a three-dimensional cavity with one oscillating boundary. Because the energy levels of the system are not equidistant, only a limited number of modes of the EM field are coupled for a specific oscillation frequency of the boundary. In addition, relevant to Fermi's golden rule, there are only three kinds of resonance conditions, which significantly simplify the calculation of the characteristic function of the joint probability distribution. We further obtain the analytical expressions in the DCE with several different time-dependent geometries (the rectangular, cylindrical and spherical cavities with one moving boundary). Analytical expressions of work distributions for an arbitrary nonequilibrium process is extremely rare. This example, which incorporates effects of quantum mechanics, quantum statistics and special relativity has pedagogical value, and deepens our understanding of the quantum trajectory work and the validity of the fluctuation theorems in nonequilibrium processes with photon creation and annihilation.

 We notice that Ref.~\cite{ja2009} also discussed the Jarzynski equality for the photon gas. However, we believe that the particles considered in Ref.~\cite{ja2009} are not real photons but some relativistic massless particles, since the wave character of light is ignored and no particle creation and annihilation occur during the force protocol.

 This paper is organized as follows: In Sec.~\ref{th}, we introduce the effective Hamiltonian of a quantized EM field in a three-dimensional cavity with a moving boundary. In Sec.~\ref{tw}, we clarifies the concept of trajectory work and the validity of the Jarzynski equality in the system. In Sec.~\ref{ch}, we obtain the analytical expressions of the characteristic functions by utilizing the matrix representation technique. And then we analyze the characteristic functions in the single-resonance and multiple-resonance conditions. We summarize this work and make conclusions in Sec.~\ref{co}.

\section{Effective Hamiltonian and resonance conditions for the EM field in a trembling cavity}
\label{th}

We begin with the quantization of the EM field confined in a 3-D rectangular cavity, expressed in the scalar Hertz potentials~\cite{he2005}. Then we derive the effective Hamiltonian, which describes the dynamics of the EM field when one of the boundaries is moving with time. Finally, we introduce some thermodynamic concepts relevant to the non-equilibrium driving process, where the photons are created or annihilated, including the two-point measurement and the characteristic function ${\mathcal G}(u,v)$. The study can be straightforwardly extended to the EM field in other geometries, such as the cylindrical and the spherical cavity (see Appendix~A).

\subsection{Hertz Potential Formalism}

It is known that the EM field and the Maxwell's equations can be formulated in terms of the scalar potential $\phi$ and the vector potential ${\mathbf A}$. Alternatively, the Hertz potentials, ${\mathbf \Pi}_{\rm e}$ and ${\mathbf \Pi}_{\rm m}$, offer an equivalent, more convenient for EM field in a cavity, formalism. In the Lorentz gauge, the relations between the two formulations can be written as follows,
\begin{eqnarray}
\phi=-\frac{1}{\varepsilon}\nabla\cdot{\mathbf\Pi}_e,\quad {\mathbf A}=\mu\frac{\partial{\mathbf\Pi}_{\rm e}}{\partial t}+\nabla\times{\mathbf \Pi}_{\rm m},
\end{eqnarray}
with $\varepsilon$ and $\mu$ being the permittivity and the permeability of the medium.

In source-free vacuum, the vector Hertz potentials become two scalar fields, i.e. ${\mathbf\Pi}_{\rm e}=\psi^{\rm TM}{\mathbf e}_z$ and ${\mathbf \Pi}_m=\psi^{\rm TE}{\mathbf e}_z$, where $\psi^{\rm TM}$ and $\psi^{\rm TE}$ are the transverse magnetic (TM) and the transverse electric (TE) field with respect to the longitudinal $z$ axis, with ${\mathbf e}_z$ being the unit vector along the $z$ axis. The Maxwell's equations in the form of the scalar Hertz potentials can be written in the following form,
\begin{eqnarray}
\left(\nabla^2-\varepsilon_0\mu_0\partial_t^2\right)\psi^{\rm TE,TM}=0,\label{eq:wave_equation}
\end{eqnarray}
with $\varepsilon_0$ and $\mu_0$ being the permittivity and the permeability of the vacuum.

Without loss of generality, we only consider the TE field in the following. The same procedure can be straightforwardly applied to the study of the TM field (see Appendix~B). Also, we hereafter drop the superscript (TE) and set $\varepsilon_0=\mu_0=\hbar=1$ for simplicity, if not explicitly stated otherwise.

The Lagrangian density of the TE field is
\begin{eqnarray}
{\mathcal L}({\mathbf r}, t)=\frac{1}{2}\left(-\dot\psi\nabla_\perp^2\dot\psi-\nabla^2\psi\nabla^2_{\perp}\psi\right).\label{eq:lagrangian_density}
\end{eqnarray}
Meanwhile, the TE field in a 3-D cavity satisfies the following boundary conditions,
\begin{eqnarray}
\left.\psi\right|_{z=0, L_z} = 0,\quad \left.\frac{\partial\psi}{\partial x}\right|_{x=0, L_x}=\left.\frac{\partial\psi}{\partial y}\right|_{y=0, L_y}=0,\label{eq:boundary_conditions}
\end{eqnarray}
where the boundaries of the cavity locate at 0 and $L_\alpha$, with $\alpha=x,y,z$.

\subsection{Quantization of the scalar Hertz Potential}

We quantize the scalar Hertz potential $\psi$ by promoting it from ordinary numbers to an operator $\hat\psi$ and imposing the following canonical commutation relations,
\begin{eqnarray}
\left[\hat\psi({\mathbf r}, t), \hat\psi({\mathbf r}', t)\right]&=&\left[\hat\pi({\mathbf r}, t), \hat\pi({\mathbf r}', t)\right]=0,\\
\left[\hat\psi({\mathbf r}, t), \hat\pi({\mathbf r}', t)\right]&=&i\delta({\mathbf r}-{\mathbf r}'),\nonumber
\end{eqnarray}
where the conjugate momentum operator $\hat\pi({\mathbf r}, t)$ is obtained from the Lagrangian density in Eq.~(\ref{eq:lagrangian_density}),
\begin{eqnarray}
\hat\pi({\mathbf r}, t)=-\nabla^2_{\perp}\frac{\partial\hat\psi({\mathbf r},t)}{\partial t}.
\end{eqnarray}

Solving the wave equation (\ref{eq:wave_equation}) under the boundary conditions (\ref{eq:boundary_conditions}), we obtain an orthonormal basis $\left\{\psi_{\mathbf k}({\mathbf r})\right\}$, where the basis state $\psi_{\mathbf k}({\mathbf r})$ is
\begin{eqnarray}
\psi_{\mathbf k}({\mathbf r})&=&\sqrt{\frac{2}{L_z}}\sin\left(\frac{k_z\pi}{L_z}z\right)\nonumber\\
&&\times\frac{2}{\sqrt{L_xL_y}}\cos\left(\frac{k_x\pi}{L_x}x\right)\cos\left(\frac{k_y\pi}{L_y}y\right),\label{eq:eigen_functions}
\end{eqnarray}
with the subscript ${\mathbf k}\equiv(k_x, k_y, k_z)\in{\mathbb N}^3$, and the eigen-frequency of the ${\mathbf k}$-th mode is $\omega_{\mathbf k}=\sqrt{\left(\frac{k_x\pi}{L_x}\right)^2+\left(\frac{k_y\pi}{L_y}\right)^2+\left(\frac{k_z\pi}{L_z}\right)^2}$.

Then we expand the filed operator $\hat\psi({\mathbf r},t)$ and its conjugate momentum operator $\hat\pi({\mathbf r},t)$ as follows,
\begin{eqnarray}
\hat\psi\left({\mathbf r}, t\right)&=&\sum_{\mathbf k}C_{\mathbf k}\psi_{\mathbf k}({\mathbf r})\hat a_{\mathbf k}e^{-i\omega_{\mathbf k}t}+{\rm h.c.},\label{eq:field_operator_expand}\\
\hat\pi({\mathbf r}, t)&=&-i\sum_{\mathbf k}C_{\mathbf k}\omega^2_{{\mathbf k}_\perp}\omega_{\mathbf k}\psi_{\mathbf k}({\mathbf r})\hat a_{\mathbf k}e^{-i\omega_{\mathbf k}t}+{\rm h.c.},\nonumber
\end{eqnarray}
where ${\mathbf k}_\perp=(k_x, k_y)$ and $\omega_{\mathbf{k}_\perp}=\sqrt{\left(\frac{k_x\pi}{L_x}\right)^2+\left(\frac{k_y\pi}{L_y}\right)^2}$, and $C_{\mathbf k}$ are the normalization constants.  Note that $\hat a_{\mathbf k}$ ($\hat a_{\mathbf k}^\dag$) is the annihilation (creation) operator for the ${\mathbf k}$-th mode, which satisfies the canonical commutation relations,
\begin{eqnarray}
\left[\hat a_{\mathbf k},\hat a_{{\mathbf k}'}\right]=0,\quad\left[\hat a_{\mathbf k},\hat a_{{\mathbf k}'}^\dag\right]=\delta_{{\mathbf k},{\mathbf k}'}.
\end{eqnarray}

Using Eqs.~(\ref{eq:lagrangian_density}) and (\ref{eq:field_operator_expand}) and setting $C_{\mathbf k}\equiv\left(\sqrt{2\omega_{\mathbf k}}\omega_{{\mathbf k}_\perp}\right)^{-1}$, we obtain the Hamiltonian of the TE field as follows~\cite{he2005},
\begin{eqnarray}
\hat H =\int d{\mathbf r}\left[\frac{\partial\hat\psi}{\partial t}\hat\pi-\hat L\right]=\sum_{\mathbf k}\frac{\omega_{\mathbf k}}{2}\left(\hat a_{\mathbf k}^\dagger\hat a_{\mathbf k}+\hat a_{\mathbf k}\hat a_{\mathbf k}^\dagger\right).
\end{eqnarray}

\subsection{Effective Hamiltonian for Driving Processes}

Now we consider that one of the boundaries of the cavity along the longitudinal $z$ direction is moving according to a prefixed time-dependent function $\lambda(t)$, i.e.  $L_z(t)\equiv\lambda(t)$.

We first define an instantaneous orthonormal basis $\{\psi_{{\mathbf k},\lambda}({\mathbf r})\}$. The basis functions $\psi_{{\mathbf k},\lambda}({\mathbf r})$ satisfy the following Helmholtz equation,
\begin{eqnarray}
\label{eq:helmholtz}
\nabla^2\psi_{{\mathbf k},\lambda}\left({\mathbf r}\right)+\omega_{{\mathbf k},\lambda}^2\psi_{{\mathbf k},\lambda}({\mathbf r})=0,
\end{eqnarray}
with the time-dependent boundary conditions~\cite{qu1970}
\begin{eqnarray}
&&\left.\psi_{{\mathbf k},\lambda}({\mathbf r})\right|_{z=0, \lambda} = 0,\\
&&\left.\frac{\partial\psi_{{\mathbf k},\lambda}({\mathbf r})}{\partial x}\right|_{x=0, L_x}=\left.\frac{\partial\psi_{{\mathbf k},\lambda}({\mathbf r})}{\partial y}\right|_{y=0, L_y}=0,\nonumber
\end{eqnarray}
where $\psi_{{\mathbf k},\lambda}({\mathbf r})$ have the same form as $\psi_{\mathbf k}({\mathbf r})$ in Eq.~(\ref{eq:eigen_functions}) except that $L_z$ is replaced by $\lambda$ and $\omega_{{\mathbf k},\lambda}=\sqrt{\left(\frac{k_x\pi}{L_x}\right)^2+\left(\frac{k_y\pi}{L_y}\right)^2+\left(\frac{k_z\pi}{\lambda}\right)^2}$. Note that both $\psi_{{\mathbf k},\lambda}({\mathbf r}$ and $\omega_{{\mathbf k},\lambda}$ depend on time when the boundary is moving with time. We then expand the field operators $\hat\psi({\mathbf r}, t)$ and $\hat\pi({\mathbf r}, t)$ with the instantaneous basis $\left\{\psi_{{\mathbf k},\lambda(t)}({\mathbf r})\right\}$,
\begin{eqnarray}
\hat\psi({\mathbf r}, t)&=&\sum_{\mathbf k}\hat Q_{\mathbf k}(t)\psi_{{\mathbf k},\lambda(t)}({\mathbf r}),\label{eq:phi_k_2}\\
\hat\pi({\mathbf r}, t)&=&\sum_{\mathbf k}\hat {\mathcal P}_{\mathbf k}(t)\psi_{{\mathbf k},\lambda(t)}({\mathbf r}).\label{eq:pi_k_2}
\end{eqnarray}

Taking time derivatives of Eqs.~(\ref{eq:phi_k_2}) and (\ref{eq:pi_k_2}), we obtain the following equations of $\hat Q_{\mathbf k}$ and $\hat {\mathcal P}_{\mathbf k}$,
\begin{eqnarray}
\frac{d\hat Q_{\mathbf k}(t)}{dt}&=&\frac{\hat {\mathcal P}_{\mathbf k}(t)}{\omega_{\mathbf{k}_\perp}^2}-\sum_{\mathbf p}\tilde g_{{\mathbf{kp}}}(t)\hat Q_{\mathbf p}(t),\label{eq:Q_motion}\\
\frac{d\hat {\mathcal P}_{\mathbf k}(t)}{dt}&=&-\omega_{\mathbf k,\lambda(t)}^2\omega_{{\mathbf k}_\perp}^2\hat Q_{\mathbf k}(t)-\sum_{\mathbf p}\tilde g_{\mathbf{kp}}(t)\hat {\mathcal P}_{\mathbf p}(t),\label{eq:P_motion}
\end{eqnarray}
where the coupling coefficients $\tilde g_\mathbf{kp}$ can be expressed as follows,
\begin{eqnarray}
\tilde g_{\mathbf{kp}}(t)&=&\int d{\mathbf r} \psi_{{\mathbf k},\lambda(t)}({\mathbf r})\frac{\partial\psi_{{\mathbf p},\lambda(t)}({\mathbf r})}{\partial t}\nonumber\\
&=&\frac{\dot \lambda(t)}{\lambda(t)}g_{\mathbf{kp}},
\end{eqnarray}
with
\begin{eqnarray}
g_{\mathbf{kp}}=\left\{\begin{array}{ll} \frac{(-1)^{k_z+p_z}2k_zp_z}{k_z^2-p_z^2}\delta_{{\mathbf k}_\perp,\mathbf{p}_\perp}, & k_z\neq p_z \\ 0, & k_z=p_z \end{array}\right..
\end{eqnarray}
treating Eqs. ($\ref{eq:Q_motion}$) and ($\ref{eq:P_motion}$) as the Heisenberg equations of motion, we obtain the following effective Hamiltonian,
\begin{eqnarray}
\hat H_{\rm eff}(t)&=&\sum_{\mathbf k}\left[\frac{\hat P^2_{\mathbf k}}{2\omega_{{\mathbf k}_\perp}^2}+\frac{1}{2}\omega_{{\mathbf k}_\perp}^2\omega_{\mathbf k,\lambda(t)}^2\hat Q^2_{\mathbf k}\right]\nonumber\\
&&-\sum_{\mathbf{kp}}{\mathcal G}_{\mathbf{kp}}(t)\hat {\mathcal P}_{\mathbf k}\hat Q_{\mathbf p}.
\end{eqnarray}

In order to move into the Fock representation, we introduce the following ladder operators,
\begin{eqnarray}
\hat a_{\mathbf k}(t)&=&\frac{\omega_{{\mathbf k}_\perp}^2\omega_{{\mathbf k},\lambda(t)}\hat Q_{\mathbf k}(t)+i\hat {\mathcal P}_{\mathbf k}(t)}{\sqrt{2\omega_{{\mathbf k},\lambda(t)}}\omega_{{\mathbf k}_\perp}},\label{eq:a_oper}\\
\hat a_{\mathbf k}^\dagger(t)&=&\frac{\omega_{{\mathbf k}_\perp}^2\omega_{{\mathbf k},\lambda(t)}\hat Q_{\mathbf k}(t)-i\hat {\mathcal P}_{\mathbf k}(t)}{\sqrt{2\omega_{{\mathbf k},\lambda(t)}}\omega_{{\mathbf k}_\perp}}.\nonumber
\end{eqnarray}
The first-order derivative of $\hat a_{\mathbf k}(t)$ can be written as follows,
\begin{eqnarray}
\frac{d}{dt}\hat a_{\mathbf k}(t)&=&-i\omega_{{\mathbf k},\lambda(t)}\hat a_{\mathbf k}(t)+2\gamma_{\mathbf k}\hat a_{\mathbf k}^\dagger(t)\nonumber\\
&&-2\sum_{\mathbf p}\left[h_\mathbf{kp}(t)\hat a_{\mathbf p}(t)+d_\mathbf{kp}(t)\hat a_{\mathbf p}^\dagger(t)\right],\label{eq:dot_ak}
\end{eqnarray}
with $\gamma_{\mathbf k}(t)=-\frac{\dot\omega_{\mathbf k,\lambda(t)}}{4\omega_{\mathbf k,\lambda(t)}}$ and
\begin{eqnarray}
h_\mathbf{kp}(t)&=&\frac{\dot\lambda(t)}{4\lambda(t)}\left(\sqrt{\frac{\omega_{\mathbf k,\lambda(t)}}{\omega_{\mathbf p,\lambda(t)}}}+\sqrt{\frac{\omega_{\mathbf p,\lambda(t)}}{\omega_{\mathbf k,\lambda(t)}}}\right)g_{\mathbf{kp}}(t),\nonumber\\
d_\mathbf{kp}(t)&=&\frac{\dot\lambda(t)}{4\lambda(t)}\left(\sqrt{\frac{\omega_{\mathbf k,\lambda(t)}}{\omega_{\mathbf p,\lambda(t)}}}-\sqrt{\frac{\omega_{\mathbf p,\lambda(t)}}{\omega_{\mathbf k,\lambda(t)}}}\right)g_{\mathbf{kp}}(t).\nonumber
\end{eqnarray}
Treating Eq.~(\ref{eq:dot_ak}) as the equation of motion, we obtain the effective Hamiltonian in terms of $\hat a_{\mathbf k}\equiv\hat a_{\mathbf k}(0)$ and $\hat a_{\mathbf k}^\dag\equiv\hat a_{\mathbf k}^\dag(0)$ as follows,
\begin{widetext}
\begin{eqnarray}
\hat H_{\rm eff}(t)&=&\sum_{\mathbf k}\omega_{\mathbf k,\lambda(t)}\left[\hat a_{\mathbf k}^\dagger\hat a_{\mathbf k}+\frac{1}{2}\right]-i\sum_{\mathbf k}\gamma_{\mathbf k}(t)\left(\hat a_{\mathbf k}^\dagger\hat a_{\mathbf k}^\dagger-\hat a_{\mathbf k}\hat a_{\mathbf k}\right)\nonumber\\
&&-i\sum_{{\mathbf k}\neq{\mathbf p}}h_{\mathbf{kp}}(t)\left(\hat a_{\mathbf k}^\dagger\hat a_{\mathbf p}-\hat a_{\mathbf p}^\dagger\hat a_{\mathbf k}\right)-i\sum_{{\mathbf k}\neq{\mathbf p}}d_{\mathbf{kp}}(t)\left(\hat a_{\mathbf k}^\dagger\hat a_{\mathbf p}^\dagger-\hat a_{\mathbf p}\hat a_{\mathbf k}\right).\label{eq:H_eff}
\end{eqnarray}
\end{widetext}
This effective Hamiltonian (\ref{eq:H_eff}) will be essential for our later analysis.

\subsection{Perturbative Driving Processes}

For simplicity, we consider the perturbative periodic driving protocol, in which the work parameter $\lambda(t)$ takes the following form,
\begin{eqnarray}
\lambda(t)=\lambda_0\left[1+\epsilon\sin(\Omega t)\right],\label{eq:lambda_t}
\end{eqnarray}
where $\Omega$ is the oscillation frequency, and the oscillation amplitude $\epsilon$ is assumed to be small, i.e. $\epsilon\ll1$. We then expand all relevant physical quantities to the first order of $\epsilon$.

In the following, we turn to the interaction picture defined by the free Hamiltonian $\hat H_0=\sum_{\mathbf k}\omega_{\mathbf k}\left(\hat a_{\mathbf k}^\dag\hat a_{\mathbf k}+\frac{1}{2}\right)$, with $\omega_{\mathbf k}\equiv\omega_{\mathbf k,\lambda(0)}$, and treat $\hat V(t)=\hat H_{\rm eff}(t)-\hat H_0$ as the perturbation. Similar to Fermi's golden rule, if $\hat V(t)$ is a periodic function of time with the angular frequency $\Omega$, the transition is into states with energies that differ by $\hbar \Omega$ from the energy of the initial state.

Using the rotating-wave approximation (RWA), we obtain time-independent Hamiltonians $\hat V_j^{\rm I}$ ($j=1,2,3$) in the interaction picture under the following three kinds of resonance conditions:
\begin{enumerate}
\item The Double-frequency (DoF) resonance: $\Omega = 2\omega_{\mathbf k}$
\begin{eqnarray}
\hat V_1^{\rm I}&=&-\frac{ig_1}{2}\left(\hat a_{\mathbf k}^\dag\hat a_{\mathbf k}^\dag-\hat a_{\mathbf k}\hat a_{\mathbf k}\right),\label{eq:V_1I}\\
g_1&=&\frac{\epsilon\Omega k_z^2\pi^2}{4\omega_{\mathbf k}^2\lambda_0^2},\nonumber
\end{eqnarray}
\item The Sum-frequency (SuF) resonance: $\Omega = \omega_{\mathbf k}+\omega_{\mathbf p}$
\begin{eqnarray}
\hat V_2^{\rm I}&=&-ig_2\left(\hat a_{\mathbf k}^\dag\hat a_{\mathbf p}^\dag-\hat a_{\mathbf p}\hat a_{\mathbf k}\right),\label{eq:V_2I}\\
g_2&=&\frac{\epsilon\Omega}{4}\left(\sqrt{\frac{\omega_{\mathbf k}}{\omega_{\mathbf p}}}-\sqrt{\frac{\omega_{\mathbf p}}{\omega_{\mathbf k}}}\right)g_\mathbf{kp},\nonumber
\end{eqnarray}
\item The Difference-frequency (DiF) resonance: $\Omega = \left|\omega_{\mathbf k}-\omega_{\mathbf p}\right|$
\begin{eqnarray}
\hat V_3^{\rm I}&=&-ig_3\left(\hat a_{\mathbf k}^\dag\hat a_{\mathbf p}-\hat a_{\mathbf p}^\dag\hat a_{\mathbf k}\right),\label{eq:V_3I}\\
g_3&=&\frac{\epsilon\Omega}{4}\left(\sqrt{\frac{\omega_{\mathbf k}}{\omega_{\mathbf p}}}+\sqrt{\frac{\omega_{\mathbf p}}{\omega_{\mathbf k}}}\right)g_\mathbf{kp}.\nonumber
\end{eqnarray}
\end{enumerate}
In addition, we can draw intuitions from Eqs.~(\ref{eq:V_1I})-(\ref{eq:V_3I}). The DoF resonance condition (\ref{eq:V_1I}) corresponds to the process of simultaneously creating two photons with the same frequency $\omega_{\mathbf k}$, while the SuF resonance condition (\ref{eq:V_2I}) corresponds to the process of creating one photon with the frequency $\omega_{\mathbf k}$ and another photon with the frequency $\omega_{\mathbf p}$. The DiF resonance condition (\ref{eq:V_3I}) corresponds to the process of creating one photon with the frequency $\omega_{\mathbf k}$ meanwhile annihilate one photon with the frequency $\omega_{\mathbf p}$.
It is worth mentioning that the evolution of other modes are quantum adiabatic with no photons being created or annihilated in these modes. Also, if the driving frequency $\Omega$ does not satisfy any of these resonance conditions (\ref{eq:V_1I})-(\ref{eq:V_3I}), the evolution of the EM field is quantum adiabatic, i.e., no photons will be created or annihilated during the driving process.

\section{Trajectory work and the Jarzynski equality associated with the DCE}
\label{tw}


We are interested in the statistics of the work and the photon number difference of the EM field confined in a 3-D cavity driven by a perturbatively oscillating boundary. Before $t=0$ and after $t=\tau$, with $\tau$ being the driving duration, the boundary stops at $\lambda_0\equiv\lambda(0)$ and $\lambda_\tau\equiv\lambda(\tau)$ respectively, while it oscillates according to Eq.~(\ref{eq:lambda_t}) when $t\in[0,\tau]$. During the driving process, both the work and the photon number fluctuate, which requires the grand-canonical description~\cite{wo2011}.

When the boundary is fixed at $\lambda$, the Hamiltonian $\hat H_\lambda=\sum_{\mathbf k}\omega_{{\mathbf k},\lambda}\hat a_{{\mathbf k},\lambda}^\dag\hat a_{{\mathbf k},\lambda}$ and the photon-number operator $\hat N_\lambda=\sum_{\mathbf k}\hat a_{{\mathbf k},\lambda}^\dag\hat a_{{\mathbf k},\lambda}$ commute with each other, i.e. $\left[\hat H_\lambda,\hat N_\lambda\right]=0$. Thus there exists a set of common eigenstates $\left|\left\{n_{\mathbf k}\right\}\right\rangle_\lambda\equiv\otimes_{\mathbf k}\ket{n_{\mathbf k}}_\lambda$, which satisfy
\begin{eqnarray}
\hat H_\lambda\ket{\set{n_{\mathbf k}}}_\lambda&=&E_\lambda\left(\set{n_{\mathbf k}}\right)\ket{\set{n_{\mathbf k}}}_\lambda,\\
\hat N_\lambda\ket{\set{n_{\mathbf k}}}_\lambda&=&N\left(\set{n_{\mathbf k}}\right)\ket{\set{n_{\mathbf k}}}_\lambda,\nonumber
\end{eqnarray}
where the total energy $E_\lambda\left(\set{n_{\mathbf k}}\right)=\sum_{\mathbf k}\omega_{{\mathbf k},\lambda} n_{\mathbf k}$ and the total photon number $N\left(\set{n_{\mathbf k}}\right)=\sum_{\mathbf k}n_{\mathbf k}$, with $n_{\mathbf k}\in{\mathbb N}$ being the number of photons in the ${\mathbf k}$-th mode. The density matrix and the partition function of the grand canonical ensemble of the photon gas is written as follows,
\begin{eqnarray}
\hat\rho_\beta&=&{\mathcal Z}_\lambda^{-1}\sum_{\set{n_{\mathbf k}}}e^{-\beta E_\lambda\left(\set{n_{\mathbf k}}\right)}\ket{\set{n_{\mathbf k}}}_\lambda{}_\lambda\langle \{n_{\mathbf k}\}|,\label{eq:rho_beta}\\
{\mathcal Z}_\lambda&=&\sum_{\set{n_{\mathbf k}}}e^{-\beta E_\lambda\left(\set{n_{\mathbf k}}\right)},\nonumber
\end{eqnarray}
with $\beta$ being the inverse temperature of the initial equilibrium state. Here we would like to emphasize that although the chemical potential of the photon gas is equal to zero, i.e. $\mu=0$, the total photon number is indefinite in thermal equilibrium.

To obtain the joint probability distribution of work and photon-number difference, ${\mathcal P}(w,\Delta N)$, the conceptual procedure of the two-point measurement protocol is prescribed as follows:
\begin{enumerate}
\item Prepare the system in the thermal equilibrium by connecting it with a heat bath at the temperature $\beta^{-1}$. Then remove the heat bath so that the system is isolated.
\item Perform the first projective measurements of $\hat H_{\lambda_0}$ and $\hat N_{\lambda_0}$. Project the system to one of the common eigenstates, i.e. $\ket{\set{n_{\mathbf k}}}_{\lambda_0}$, and record the eigenvalues $E_{\lambda_0}\left(\set{n_{\mathbf k}}\right)$ and $N\left(\set{n_{\mathbf k}}\right)$.
\item Control the boundary according to Eq.~(\ref{eq:lambda_t}) for a prefixed duration $\tau$. The frequency $\Omega$ is chosen such that one of the resonance conditions (\ref{eq:V_1I}--\ref{eq:V_3I}) is satisified,
\item Perform the second projective measurements of $\hat H_{\lambda_\tau}$ and $\hat N_{\lambda_\tau}$ and record the eigenvalues $E_{\lambda_\tau}\left(\set{n'_{\mathbf k}}\right)$ and $N\left(\set{n'_{\mathbf k}}\right)$.
\end{enumerate}
Ideally, the above procedure is repeated infinitely many times to obtain a good statistics of the joint probability distribution, which is defined as
\begin{widetext}
\begin{eqnarray}
{\mathcal P}(w,\Delta N)&=&{\mathcal Z}_{\lambda_0}^{-1}\sum_{\set{n_{\mathbf k}},\set{n'_{\mathbf k}}}e^{-\beta E_{\lambda_0}\left(\set{n_{\mathbf k}}\right)}\left|\tensor[_{\lambda_\tau}]{\evol{\set{n'_{\mathbf k}}}{\hat U(\tau)}{\set{n_{\mathbf k}}}}{_{\lambda_0}}\right|^2\nonumber\\
&&\times\delta\left(w-\left[E_{\lambda_\tau}\left(\set{n'_{\mathbf k}}\right)-E_{\lambda_0}\left(\set{n_{\mathbf k}}\right)\right]\right)\delta_{\Delta N,N\left(\set{n'_{\mathbf k}}\right)-N\left(\set{n_{\mathbf k}}\right)},\label{eq:joint_dist}
\end{eqnarray}
\end{widetext}
with $\delta(\cdot)$ and $\delta_{\cdot,\cdot}$ being the Dirac and the Kronecker delta functions, respectively.

Alternatively, we can calculate the characteristic function of the work and the photon-number difference, which is defined as the Fourier transform of ${\mathcal P}(w,\Delta N)$,
\begin{eqnarray}
{\mathcal G}(u,v)&=&\sum_{\Delta N}\int dw e^{iuw+iv\Delta N}P\left(w,\Delta N\right)\label{eq:characteristic_function}\\
&=&\avg{e^{iu\hat H^{\rm H}_{\lambda_\tau}(\tau)+iv\hat N^{\rm H}_{\lambda_\tau}(\tau)}e^{-iu\hat H_{\lambda_0}-iv\hat N_{\lambda_0}}}_\beta,\nonumber
\end{eqnarray}
where the Heisenberg-picture operators are defined as $\hat O^{\rm H}(t)=\hat U^\dag(t)\hat O\hat U(t)$ and the ensemble average is taken over $\hat\rho_\beta$ in Eq.~(\ref{eq:rho_beta}), i.e. $\avg{\cdot}_\beta={\rm Tr}\left[\cdot\hat\rho_\beta\right]$. Then the joint probability distribution ${\mathcal P}(w,\Delta N)$ can be obtained via the inverse Fourier transform.

Although there exists no Sch\"{o}rdinger picture description in our system during the driving process~\cite{qu1970}, the Hamiltonian and the total photon number operators in the Heisenberg picture before and after the driving process are well-defined. Especially at the end of the driving process, they are defined as follows,
\begin{eqnarray}
\hat H_{\lambda_\tau}^{\rm H}(\tau)&=&\sum_{\mathbf k}\omega_{\mathbf k,\lambda_\tau}\left(\hat a_{\mathbf k}^\dag(\tau)\hat a_{\mathbf k}(\tau)+\frac{1}{2}\right),\label{eq:ops_H}\\
\hat N_{\lambda_\tau}^{\rm H}(\tau)&=&\sum_{\mathbf k}\left(\hat a_{\mathbf k}^\dag(\tau)\hat a_{\mathbf k}(\tau)+\frac{1}{2}\right),\nonumber
\end{eqnarray}
where the time-dependent ladder operators $\hat a_{\mathbf k}(t)$ at the beginning ($t=0$) and the end ($t=\tau$) of the driving process are related via the unitary transformation defined by $\hat H_{\rm eff}$ in Eq.~(\ref{eq:H_eff}),
\begin{eqnarray}
\hat{a}_{\mathbf k}(\tau)=e^{i\hat{V}^{\rm I}\tau}e^{i\hat{H}_0 \tau}\hat{a}_{\mathbf k}e^{-i\hat{H}_0 \tau}e^{-i\hat{V}^{\rm I}\tau},\label{eq:lad_oper}
\end{eqnarray}
with $V^{\rm I}$ being one of $V_j^{\rm I}$ in Eqs.~(\ref{eq:V_1I}--\ref{eq:V_3I}), depending on the type of the resonance condition satisfied in the driving process.

The Jarzynski equality for a grand canonical ensemble,
 \be\label{eq:jarzynski}
 \langle e^{-\beta (w-\mu \Delta N)}\rangle=e^{-\beta\Delta\Phi},
 \ee
is obtained by setting $u=i\beta$ and $v=-i\beta\mu$, where we have defined the grand potential difference $\Delta\Phi=\Phi_{\lambda_\tau}-\Phi_{\lambda_0}$ with the grand potential $\Phi_{\lambda_{\tau,0}}=-\beta^{-1}\mathrm{ln}{\rm Tr}\left[e^{-\beta\left(\hat {H}_{\lambda_\tau}-\mu\hat {N}_{\lambda_\tau}\right)}\right]$.

The equality in Eq.~(\ref{eq:jarzynski}) reproduces the original Jarzynski equality, $\langle e^{-\beta w}\rangle=e^{-\beta\Delta\Phi}$, in the following two cases
 \begin{subequations}
 \bey
 \label{e24}
 (i) &\mu\neq0,\quad\hat{N}_{\lambda_\tau}^{\rm H}(\tau)=\hat{N}_{\lambda_0},\\
 \label{e25}
 (ii) &\mu=0,\quad\hat{N}_{\lambda_\tau}^{\rm H}(\tau)\neq\hat{N}_{\lambda_0}.
 \eey
 \end{subequations}
 We would like to emphasize that Refs.~\cite{no2011,wo2011,wo2012} have discussed systems that belong to case (i), while our system belongs to case (ii).

\section{analytical results of work distributions under various resonance conditions}
\label{ch}

Generally, it is impossible to obtain closed-form expressions of ${\mathcal G}(u,v)$ for an arbitrary non-equilibrium driving processes of a quantum many-body system, let alone the case with particle creation and annihilation. However, we find that for the photon gas confined in a cavity subject to a perturbative resonant driving (\ref{eq:lambda_t}), the characteristic function ${\mathcal G}(u,v)$ can be written into the expectation value of a product of a series of exponentials of a quadratic form in the Boson operators. By substituting Eqs.~(\ref{eq:ops_H}) and (\ref{eq:lad_oper}) into Eq.~(\ref{eq:characteristic_function}) and then by applying the matrix representation technique~\cite{en1978}, we obtain the analytical results of the characteristic function ${\mathcal G}(u,v)$ (for more technical details see Appendix~C).

\subsection{Characteristic Functions in Single-Resonance Conditions}

\begin{figure*}
\includegraphics[width=\textwidth]{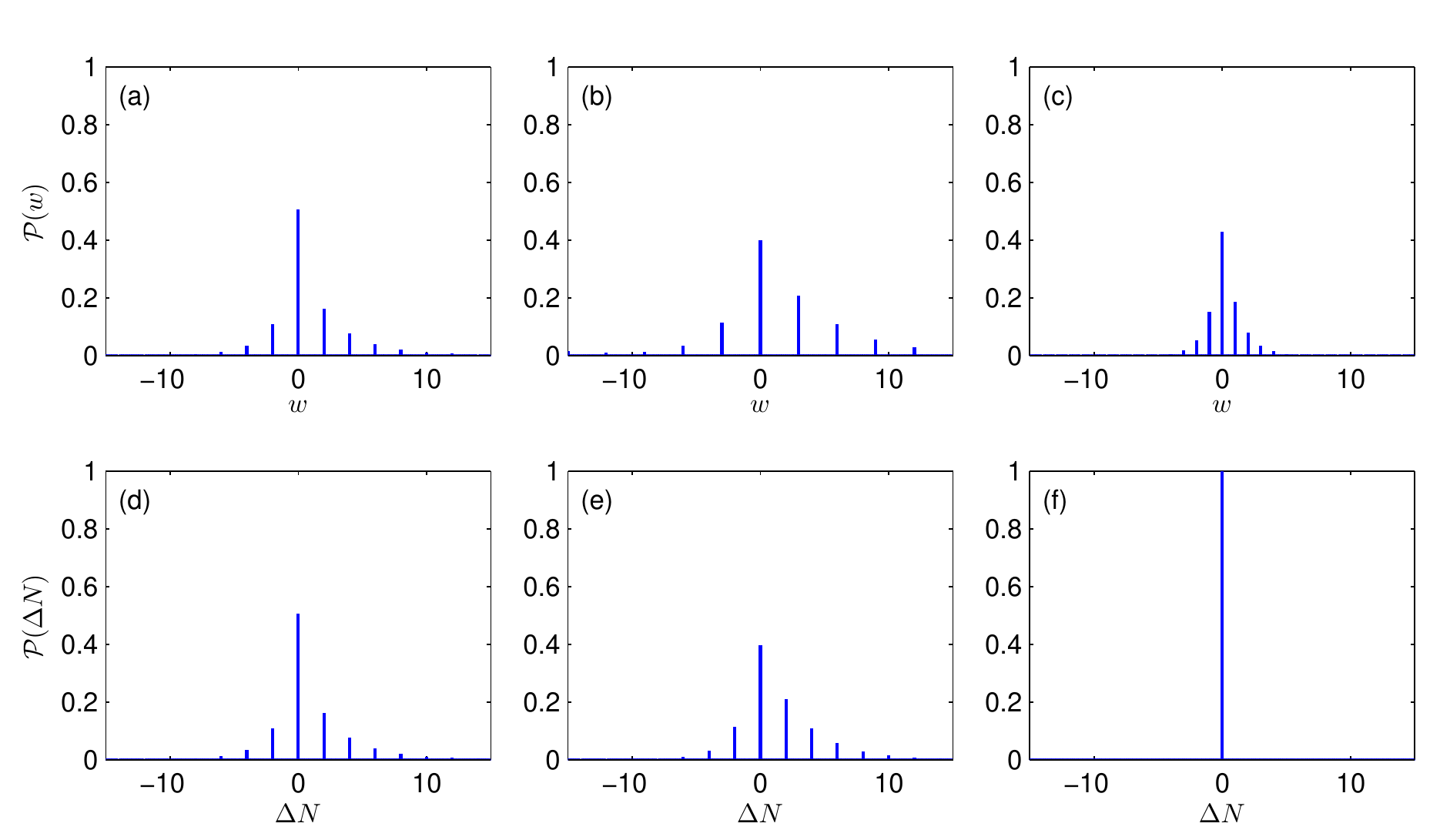}
\caption{The marginal distributions of work ${\mathcal P}(w)$ (a--c) and the photon number change ${\mathcal P}(\Delta N)$ (d--f). The columns from the left to the right are for the cases of the DoF, the SuF and the DiF resonance conditions, respectively. The parameters are chosen as follows: $\beta\omega_{\mathbf k}=0.2$, $g_j\tau=0.3$ ($j=1,2,3$), and $\omega_{\mathbf p}=2\omega_{\mathbf k}$.}
\label{fig:marginal_distributions}
\end{figure*}

If $\Omega$ is chosen such that one of the three resonance conditions (\ref{eq:V_1I})-(\ref{eq:V_3I}) is satisfied, the perturbation Hamiltonian becomes independent of time (see Eqs.~(\ref{eq:V_1I}--\ref{eq:V_3I})). In these cases, we can analytically calculate the characteristic function ${\mathcal G}(u,v)$ in Eq.~(\ref{eq:characteristic_function}). The results are given below:
\begin{widetext}
\begin{enumerate}
\item The DoF resonance: $\Omega = 2\omega_{\mathbf k}$
\begin{eqnarray}\label{eq:closed_form_G1}
{\mathcal G}_1(u,v)=\frac{\sinh\frac{\beta\omega_{\mathbf k}}{2}}{\sqrt{\sinh^2\frac{\beta\omega_{\mathbf k}}{2}+\sin\left(u\omega_{\mathbf k}+v\right)\sin\left(\left(u-i\beta\right)\omega_{\mathbf k}+v\right)\sinh^2g_1\tau}},
\end{eqnarray}
\item The SuF resonance: $\Omega = \omega_{\mathbf k}+\omega_{\mathbf p}$
\begin{eqnarray}\label{eq:closed_form_G2}
{\mathcal G}_2(u,v)=\frac{\sinh\frac{\beta\omega_{\mathbf k}}{2}\sinh\frac{\beta\omega_{\mathbf p}}{2}}{\sinh\frac{\beta\omega_{\mathbf k}}{2}\sinh\frac{\beta\omega_{\mathbf p}}{2}+\sin\left(\frac{u\left(\omega_{\mathbf k}+\omega_{\mathbf p}\right)}{2}+v\right)\sin\left(\frac{\left(u-i\beta\right)\left(\omega_{\mathbf k}+\omega_{\mathbf p}\right)}{2}+v\right)\sinh^2g_2\tau},
\end{eqnarray}
\item The DiF resonance: $\Omega = \left|\omega_{\mathbf k}-\omega_{\mathbf p}\right|$
\begin{eqnarray}\label{eq:closed_form_G3}
{\mathcal G}_3(u,v)=\frac{\sinh\frac{\beta\omega_{\mathbf k}}{2}\sinh\frac{\beta\omega_{\mathbf p}}{2}}{\sinh\frac{\beta\omega_{\mathbf k}}{2}\sinh\frac{\beta\omega_{\mathbf p}}{2}+\sin\left(\frac{u\left(\omega_{\mathbf k}-\omega_{\mathbf p}\right)}{2}\right)\sin\left(\frac{\left(u-i\beta\right)\left(\omega_{\mathbf k}-\omega_{\mathbf p}\right)}{2}\right)\sin^2g_3\tau}.
\end{eqnarray}
\end{enumerate}
\end{widetext}
Note that we restrict ourselves to the cases with $\lambda_0=\lambda_\tau$ for simplicity. The analytic solutions for more general cases with $\lambda_0\neq\lambda_\tau$ are more cumbersome and can be found in Appendix~D.

Note that the characteristic function ${\mathcal G}(u,v)$ for an externally-driven closed quantum system should satisfy the following requirements:
\begin{enumerate}
\item The normalization of the joint probability ${\mathcal P}(w,\Delta N)$, i.e. $\sum_{\Delta N}\int dw {\mathcal P}(w,\Delta N)=1$, requires that ${\mathcal G}(0, 0)=1$.
\item The grand-canonical quantum Jarzynski equality, i.e. $\avg{e^{-\beta\left(w-\mu\Delta N\right)}}=e^{-\beta\Delta\Phi}$, 
    requires that ${\mathcal G}(i\beta,-i\beta\mu)=e^{-\beta\Delta\Phi}$.
\item The grand-canonical Crooks' fluctuation theorem, i.e. $\frac{{\mathcal P}_{\rm F}\left(w,\Delta N\right)}{{\mathcal P}_{\rm R}\left(-w,-\Delta N\right)}=e^{\beta\left(w-\nu\Delta N-\Delta\Phi\right)}$, where the forward and the reverse processes are respectively described by $\lambda(t)$ and $\lambda(\tau-t)$, requires that ${\mathcal G}_{\rm R}(-u,-v)={\mathcal G}_{\rm F}(u+i\beta,v-i\beta\mu)e^{\beta\Delta\Phi}$.
\item The discreteness of the distribution function of work for a closed quantum system requires that ${\mathcal G}(u,v)$ is periodic in $u$. More specifically, a delta-function peak $\delta(w-w_0)$ in ${\mathcal P}(w)$ implies that ${\mathcal G}\left(u+\frac{2\pi}{\omega_0},v\right)={\mathcal G}(u,v)$.
\item The indivisibility of the constituent photons, $\Delta N\in{\mathbb Z}$, requires that ${\mathcal G}(u,v+2\pi)={\mathcal G}(u,v)$. Furthermore, if only $n>0$ photons can be created or annihilated simultaneously, ${\mathcal G}\left(u,v+\frac{2\pi}{n}\right)={\mathcal G}(u,v)$. Finally, if the photon creation and annihilation processes are prohibited, ${\mathcal G}(u,v)$ is independent of $v$, i.e. ${\mathcal G}(u,v)={\mathcal G}(u)$.
\end{enumerate}

For the particular situation we are interested in, the EM field confined in a 3-D cavity resonantly driven by a trembling boundary, the chemical potential vanishes and the positions of the trembling boundary at the initial and final time are exactly the same. It can be checked that Eqs.~(\ref{eq:closed_form_G1}--\ref{eq:closed_form_G3}) satisfy all the above requirements.

\subsection{Marginal Distributions of the work and the photon-number difference}
The joint distribution function ${\mathcal P}(w,\Delta N)$ can be obtained from the inverse Fourier transform of the characteristic function ${\mathcal G}(u,v)$,
\begin{eqnarray}
{\mathcal P}(w,\Delta N)=\frac{1}{4\pi^2}\int du\int dv e^{-iuw-iv\Delta N}{\mathcal G}(u,v),
\end{eqnarray}
from which we obtain the following marginal distributions ${\mathcal P}(w)$ and ${\mathcal P}(\Delta N)$,
\begin{eqnarray}
{\mathcal P}(w)&=&\sum_{\Delta N}{\mathcal P}(w,\Delta N)\nonumber\\
&=&\frac{1}{2\pi}\int du e^{-iuw} {\mathcal G}(u,0),\label{eq:marginal_PW}\\
{\mathcal P}(\Delta N)&=&\int dw {\mathcal P}(w,\Delta N)\nonumber\\
&=&\frac{1}{2\pi}\int dv e^{-iv\Delta N} {\mathcal G}(0,v).\label{eq:marginal_PN}
\end{eqnarray}
In Fig.~\ref{fig:marginal_distributions}, we show the marginal distributions for all of the three kinds of resonance conditions. Since we are considering closed systems with discrete energy levels, all of the distributions are discrete, consisting of a series of the Dirac delta-functions with different weights. The work performed on the system is always an integer multiple of $2\omega_{\mathbf k}$, $\omega_{\mathbf k}+\omega_{\mathbf p}$ or $\omega_{\mathbf k}-\omega_{\mathbf p}$, when the DoF (\ref{eq:V_1I}), the SuF (\ref{eq:V_2I}) or the DiF (\ref{eq:V_3I}) resonance condition is satisfied, respectively. As to the photon number difference $\Delta N$, the photons are created and annihilated in pairs in the first two cases. And the numbers of the photons created and annihilated are equal to each other in the DiF resonance condition~\cite{footnote1}.  These features are consistent with our understandings of the three resonance conditions in Eqs.~(\ref{eq:V_1I}--\ref{eq:V_3I}).

\begin{figure}
\includegraphics[width=0.5\textwidth]{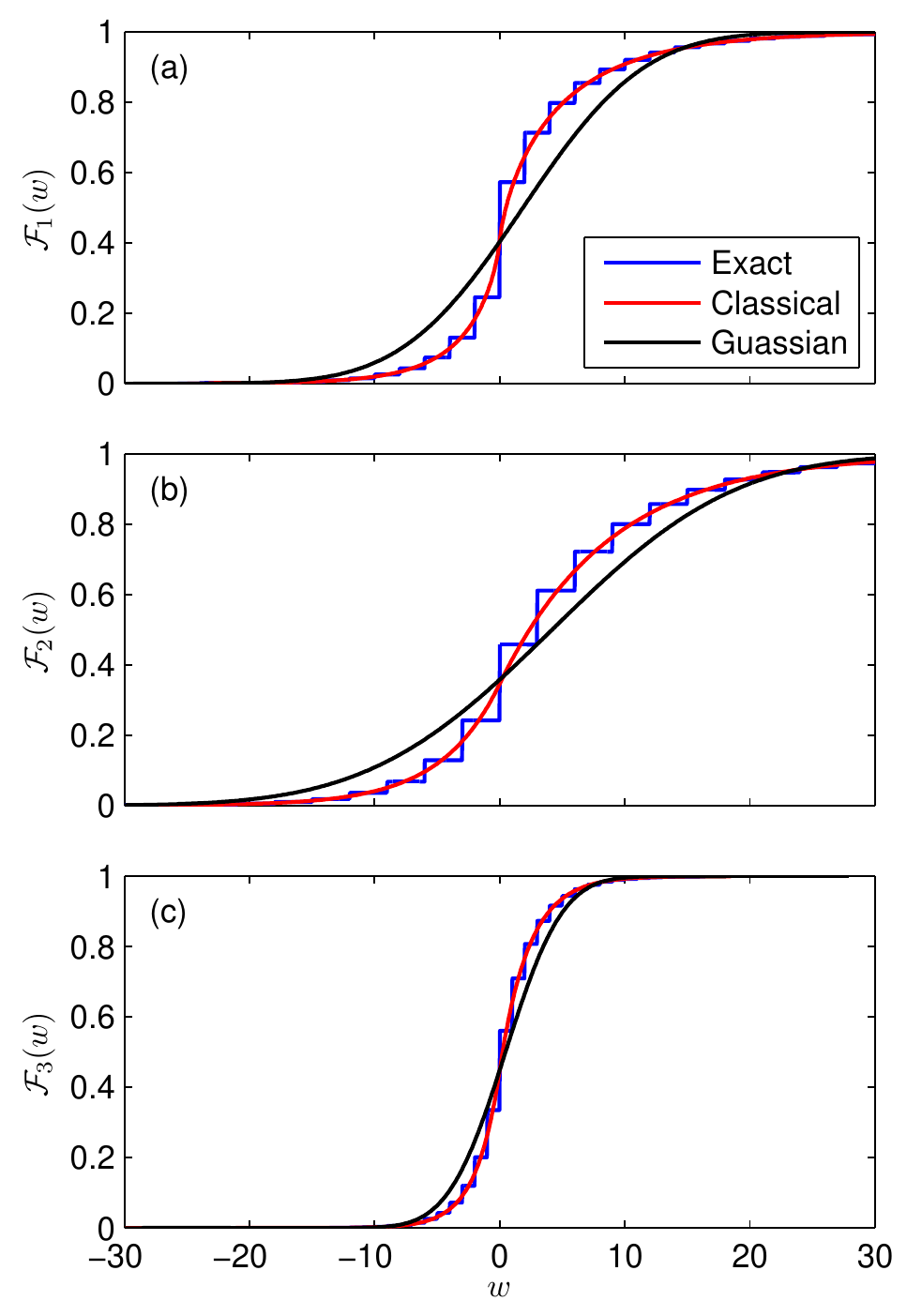}
\caption{The cumulative distributions of work. Here ${\mathcal F}_j(w)= \int_0^{w}{\mathcal P}_j(w')dw'$. The panels from the top to the bottom are for the DoF (a), the SuF (b), and the DiF (c) resonance conditions respectively. The parameters are chosen as follows: $\beta\omega_{\mathbf k}=0.1$, $g_j\tau=0.3$ with $j=1,2,3$, and $\omega_{\mathbf k}=2\omega_{\mathbf p}$ for the latter two cases. The blue step-wise lines are the exact cumulative distributions obtained from the inverse Fourier transform of $\mathcal G_j(u,0)$ while the red lines are the semiclassical ones and the black lines are the cumulative Gaussian fitting with the same average value and the standard deviation.}\label{fig:semiclassical}
\end{figure}

\subsection{Classical Limit of the Work Distributions}

The classical characteristic functions can be obtained from the closed-form expressions in Eqs.~(\ref{eq:closed_form_G1}--\ref{eq:closed_form_G3}) by taking the classical limit, i.e. $\hbar\rightarrow0$~\cite{footnote3}. Similar to Ref.~\cite{wo2011}, we introduce $\tilde u=u/\beta$.

The classical characteristic functions for the three resonance conditions can be written as follows~\cite{footnote5},
\begin{eqnarray}
{\mathcal G}_1^{\rm cl}(\tilde u,0)&=&\frac{1}{\sqrt{1+4(\tilde u^2-i\tilde u)\sinh^2g_1\tau}},\\
{\mathcal G}_2^{\rm cl}(\tilde u,0)&=&\frac{1}{1+\frac{(r+1)^2}{r}\left(\tilde u^2-i\tilde u\right)\sinh^2g_2\tau},\nonumber\\
{\mathcal G}_3^{\rm cl}(\tilde u,0)&=&\frac{1}{1+\frac{(r-1)^2}{r}\left(\tilde u^2-i\tilde u\right)\sin^2g_3\tau},\nonumber
\end{eqnarray}
with the ratio being defined as $r\equiv\omega_{\mathbf p}/\omega_{\mathbf k}$ in the latter two cases. Note that the classical work distributions for the latter two cases can be obtained analytically as follows,
\begin{eqnarray}
{\mathcal P}_j^{\rm cl}(w)=\frac{\alpha_{j+}\alpha_{j-}}{\alpha_{j+}-\alpha_{j-}}\left[e^{\alpha_{j+}w}\Theta(w)+e^{\alpha_{j-}w}\Theta(-w)\right],
\end{eqnarray}
with $j=2, 3$, where $\Theta(\cdot)$ is the Heaviside step function and
\begin{eqnarray}
\alpha_{2\pm}&=&\frac{\beta}{2}\left(1\mp\sqrt{1+\frac{4 r{\rm csch}^2g_2\tau}{\left(r+1\right)^2}}\right),\\
\alpha_{3\pm}&=&\frac{\beta}{2}\left(1\mp\sqrt{1+\frac{4 r{\rm csc}^2g_3\tau}{\left(r-1\right)^2}}\right).\nonumber
\end{eqnarray}
Fig.~\ref{fig:semiclassical} shows the consistency between the semiclassical work distributions and the exact ones when the temperature of the initial state is high.
Also, we show the discrepancy between the exact marginal distributions of work and the Gaussian fitting with the same average value and the same standard deviation of the work distribution. It can be seen that the work distributions obviously deviate from the Gaussian distribution.

\subsection{Quantum-to-Classical Crossover}

\begin{figure*}
\includegraphics[width=\textwidth]{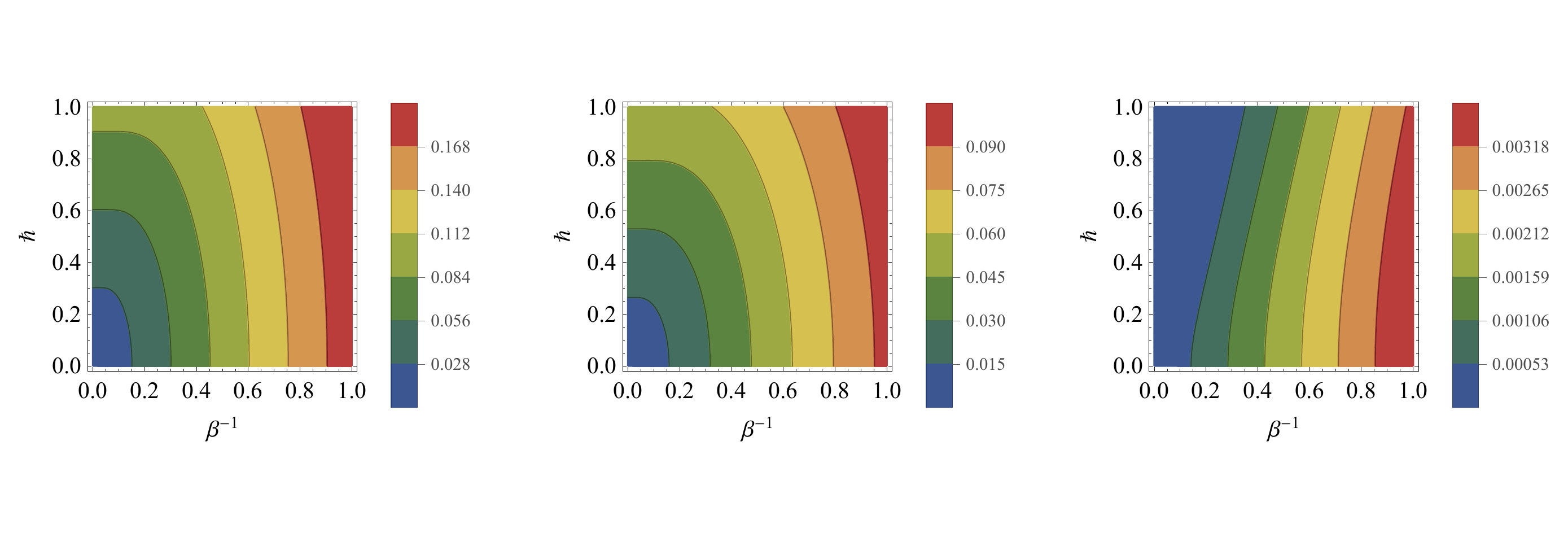}
\caption{Average work, $\avg{w}_j$ with $j=1,2,3$, for the driving protocols satisfying the DoF (left), the SuF (middle) and the DiF (right) resonance conditions, respectively. The parameters are chosen as follows: $\omega_{\mathbf k}=1$, $g_j\tau=0.3$ with $j=1,2,3$ and $r=1.5$ for the latter two cases.}\label{fig:quantumclassical}
\end{figure*}

\begin{table*}[ht]
 \centering
 \caption{Moments of work in two limits of the initial temperature. In the low temperature limit, the nonzero works $\avg{w}_1$ and $\avg{w}_2$ are manifestations of the quantum nature of the EM field (zero point energy). When $\hbar \to 0$, the works $\avg{w}_1$ and $\avg{w}_2$ become zero, which agrees with our intuition that no work is done on the classical vacuum. }
 \label{tb:moments}
 \renewcommand{\arraystretch}{2}
 \begin{tabular}{|c|c|c|}
  \hline
  \hline
  & Low-temperature limit ($\beta\rightarrow\infty$) & High-temperature limit ($\beta\rightarrow0$)\\
  \hline
  $\avg{w}_1$ & $\hbar\omega\sinh^2g_1\tau$ & $2\beta^{-1}\sinh^2g_1\tau$ \\
  \hline
  $\sigma_{w,1}^2$ & $\frac{1}{2}\left(\hbar\omega\right)^2\sinh^22g_1\tau$ & $4\beta^{-2}\cosh2g_1\tau_1\sinh^2g_1\tau$ \\
  \hline
  $\avg{w}_2$ & $(1+r)\hbar\omega\sinh^2g_2\tau$ & $\beta^{-1}\frac{(1+r)^2}{r}\sinh^2g_2\tau$ \\
  \hline
  $\sigma_{w,2}^2$ & $\frac{(1+r)^2}{4}(\hbar\omega)^2\sinh^22g_2\tau$ & $\beta^{-2}\frac{(1+r)^4}{r^2}\left(\sinh^2g_2\tau_2+\frac{2r}{(1+r)^2}\right)\sinh^2g_2\tau$ \\
  \hline
  $\avg{w}_3$ & 0 & $\beta^{-1}\frac{(r-1)^2}{r}\sin^2g_3\tau$ \\
  \hline
  $\sigma_{w,3}^2$ & 0 & $\beta^{-2}\frac{(1-r)^4}{r^2}\left(\sin^2g_3\tau_3+\frac{2r}{(1-r)^2}\right)\sin^2g_3\tau$ \\
  \hline
 \end{tabular}
 \label{tab:myfirsttable}
\end{table*}

The analytic expressions of the average value and the standard deviation of the work distributions are defined as
 \be
 \avg{w}_j\equiv-i\left.\frac{\partial \mathcal G_j(u,0)}{\partial u}\right|_{u=0},j=1,2,3,
 \ee
 \be
 \sigma^2_{w,j}\equiv-\left.\frac{\partial^2 \mathcal G_j(u,0)}{\partial^2 u}\right|_{u=0},j=1,2,3,
 \ee
 which can be obtained from Eqs.~(\ref{eq:closed_form_G1}--\ref{eq:closed_form_G3}). For the convenience of analysis, we explicitly write down the Planck constant $\hbar$. It can be anticipated that when $\hbar\rightarrow0$, the system loses all of its quantum features~\cite{footnote2}.

In Fig.~\ref{fig:quantumclassical}, we demonstrate the average work performed on the system in these three resonance conditions. Inspecting the figures, we find that the average work decreases to zero as $\hbar$ vanishes in the low-temperature limit for the DoF and the SuF cases, which is consistent with the fact that no work can be done on the classical vacuum. Also, the dependence of $\avg{w}_j$ on $\hbar$ becomes weaker for all three resonance conditions as the temperature increases, which implies a crossover from the quantum to the classical regime.

Inspired by the observations in Fig.~\ref{fig:quantumclassical}, we list in TABLE~\ref{tb:moments} the average value and the standard deviation of the work distributions for the three resonance conditions in two limits, i.e. the low-temperature ($\beta\rightarrow\infty$) and the high-temperature ($\beta\rightarrow0$) limits. It can be seen clearly that the energy scale is $\hbar\omega$ and $\beta^{-1}$ in the low-temperature and the high-temperature limits, respectively. In the high temperature limit, the work is equal to the work done on a classical EM field, which can be seen from the fact that the work value does not depend on $\hbar$.

\subsection{Multiple-Resonance cases}

If $\Omega$ is chosen such that for all modes in the cavity field, at least two resonance conditions are satisfied simultaneously,
we are dealing with a ``multiple-resonance'' condition. The multiple-resonance condition can be double-resonant, triple-resonant, and so on.

For the multiple-resonance condition, we consider the following two cases: the uncoupled and the coupled. For the uncoupled case, the involved modes are different, e.g. $\Omega=2\omega_{\mathbf{k}}=\omega_{\mathbf{s}}-\omega_{\mathbf{p}}$ with $\mathbf{k}\neq\mathbf{s}\neq\mathbf{p}$. Because the resonance conditions are mutually independent, the characteristic function can be decomposed into a product of characteristic functions, each of which can be obtained analytically in the corresponding resonance condition. Let us denote the characteristic function for a multiple-resonance case as ${\mathcal G}_{\mathbf{k}_1,\mathbf{k}_2,\cdots,\mathbf{s}_1\pm\mathbf{p}_1,\mathbf{s}_2\pm\mathbf{p}_2,\cdots}(u, v)$ for resonance conditions $\Omega=2\omega_{\mathbf{k}_1}=2\omega_{\mathbf{k}_2}=\cdots=\omega_{\mathbf{s}_1}\pm\omega_{\mathbf{p}_1}=\omega_{\mathbf{s}_2}\pm\omega_{\mathbf{p}_2}=\cdots$. The characteristic function for this case reads
\begin{eqnarray}
\label{e43}
&&{\mathcal G}_{\mathbf{k}_1,\mathbf{k}_2,\cdots,\mathbf{s}_1\pm\mathbf{p}_1,\mathbf{s}_2\pm\mathbf{p}_2,\cdots}(u, v)\\
&=&{\mathcal G}_{\mathbf{k}_1}(u, v){\mathcal G}_{\mathbf{k}_2}(u, v)\ldots\nonumber\\
&&\times {\mathcal G}_{\mathbf{s}_1\pm\mathbf{p}_1}(u, v){\mathcal G}_{\mathbf{s}_2\pm\mathbf{p}_2}(u, v)\ldots,\nonumber
\end{eqnarray}
and the joint probability distribution function ${\mathcal P}(w, \Delta N)$ can be obtained by the inverse Fourier transform.

For the coupled case, more than one resonance conditions involve the same modes, for example $\Omega=2\omega_{\mathbf{k}}=\left|\omega_{\mathbf{k}}-\omega_{\mathbf{p}}\right|$. Since the mode $\mathbf{k}$ appears in both resonance conditions,  these two resonance conditions cannot be considered separately. Instead, the perturbation Hamiltonian in the interaction picture becomes $\hat{V}^{\mathrm I}=-ig_1(\hat{a}_{\mathbf{k}}^{\dag}\hat a_{\mathbf k}^\dag-\hat{a}_{\mathbf{k}}\hat a_{\mathbf k})
 -ig_3(\hat{a}_{\mathbf{k}}^{\dag}\hat{a}_{\mathbf{p}}-\hat{a}_{\mathbf{p}}^{\dag}\hat{a}_{\mathbf{k}})]$, with which the characteristic function ${\mathcal G}(u,v)$ can also be obtained by the matrix representation technique. The discussions can be straightforwardly extended to the multiple-resonance case with more than two resonance conditions are satisfied simultaneously.

 \section{Conclusions}
 \label{co}

To calculate the work distribution in an arbitrary nonequilibrium process in a quantum many-body system is usually very cumbersome due to the interplay of effects of quantum mechanics and quantum statistics \cite{gong2014}. In very rare situations, one is able to obtain the analytical solution to the distribution of work. These analytical results deepens our understanding of quantunm trajectory work and fluctuation theorems in the nonequilibrium processes.

In our current study, we investigated the work distribution of a quantized EM field in a three-dimensional cavity with an oscillating boundary. This system incorporates the effects of not only quantum mechanics and quantum statistics, but also special relativity. For the periodic perturbative driving protocol (\ref{eq:lambda_t}), under the RWA, we obtained the effective Hamiltonian in the interaction picture. Also, we analytically evaluate the characteristic function ${\mathcal G}(u,v)$ in the single-resonance (\ref{eq:closed_form_G1}--\ref{eq:closed_form_G3})  and multiple-resonance conditions using the matrix representation technique. If $\Omega$ is chosen such that none of the single resonance conditions is satisified, the evolution of the EM field is quantum adiabatic. We discussed the general properties of ${\mathcal G}(u,v)$ and verified various fluctuation theorems in the nonequilibrium processes with photon creation and annihilation. From the analytical result of the work distribution, we can clearly see that nonzero work is done at zero temperature, which is a manifestation of the quantum nature (Casimir effect) of the EM field. However, the work vanishes when $\hbar \to 0$, which agrees with our intuition that no work is done on the classical vacuum when the boundary is driven. We also obtained the approximate expression of the work distribution ${\mathcal P}(w)$ and the moments of work at high temperature, which is consistent with the work done on a classical EM field. Our study has pedagogical value because analytical solutions to the work distribution in a quantum many-body system is very rare.
Last but not least, the dynamical Casimir effect has been experimentally tested in a superconducting circuit \cite{johansson2010,ob2011}. Hopefully, our theoretical predictions about the work distributions and the validity of the Jarzynski equality can be experimentally tested in the superconducting circuit~\cite{ob2011} in the future.

\begin{acknowledgments}
H. T. Quan gratefully acknowledges support from
the National Science Foundation of China under grants
11775001, 11534002, and The Recruitment Program of
Global Youth Experts of China. Jing-Ning Zhang gratefully acknowledges the support from the National Natural Science Foundation of China (Grants No. 11504197).
\end{acknowledgments}

 \renewcommand{\theequation}{A.\arabic{equation}}

 \setcounter{equation}{0}

 \appendix

 \section*{Appendix A: EM field in various geometries}
 \label{Ab}

 Our results of the EM field in a 3-D trembling rectangular cavity can be generalized to the EM field in a trembling cylindrical or spherical cavity. We skip the derivation which is similar to that of the rectangular cavity and list the main results here. The work parameter $\lambda(t)$ takes the form $\lambda(t)=\lambda_0[1+\epsilon\mathrm{sin}(\Omega t)]$ in the following, and we take the first-order approximation when doing expansion in $\epsilon$.

 In a cylindrical cavity, we introduce the cylindrical coordinates $(\rho, \phi, z)$. If the boundaries are at the radial coordinate $\rho=R$ and the longitudinal coordinate $z=(0,\lambda(t))$, we have

\begin{enumerate}
 \item Hertz potentials
 \begin{gather}
  \begin{split}
 \mathbf{\Pi}_e=\psi^{\mathrm{TM}}\hat{\mathbf{e}}_z,
 \ \ \ \ \ \mathbf{\Pi}_m=\psi^{\mathrm{TE}}\hat{\mathbf{e}}_z,
  \end{split}
 \end{gather}
 \item Boundary conditions
 \be
 \left.\psi^{\mathrm{TE}}\right|_{z=0,\lambda(t)}=0,\ \ \ \ \left.\partial_{\rho}\psi^{\mathrm{TE}}\right|_{\rho=R}=0,
 \ee
 \begin{gather}
  \begin{split}
  \label{e21}
 &\left.\partial_{z}\psi^{\mathrm{TM}}\right|_{z=0}=0,\ \left.(\partial_{z}+\dot{\lambda}(t)\partial_t)\psi^{\mathrm{TM}}\right|_{z=L_z(t)}=0,\\
 &\left.\psi^{\mathrm{TM}}\right|_{\rho=R}=0,
  \end{split}
 \end{gather}
 \item Instantaneous orthonormal bases
 \begin{gather}
  \begin{split}
 \psi^{\mathrm{TE}}_{nmk,\lambda(t)}(\mathbf{r})=&\sqrt{\frac{2}{\lambda(t)}}\mathrm{sin}(\frac{\pi kz}{\lambda(t)})\\
                                               &\times\frac{\sqrt{2}J_n(y_{nm}\rho/R)}{RJ_{n}(y_{nm})\sqrt{1-n^2/y^2_{nm}}}\frac{e^{in\phi}}{\sqrt{2\pi}},
  \end{split}
 \end{gather}
 \begin{gather}
  \begin{split}
 \psi^{\mathrm{TM}}_{nmk,\lambda(t)}(\mathbf{r})=&\sqrt{\frac{2}{\lambda(t)}}\mathrm{cos}(\frac{\pi kz}{\lambda(t)})\ \ \ \ \ \ \ \ \ \ \ \ \ \ \ \ \ \ \ \ \ \\
                                               &\times\frac{\sqrt{2}J_n(x_{nm}\rho/R)}{RJ_{n+1}(x_{nm})}\frac{e^{in\phi}}{\sqrt{2\pi}},
  \end{split}
 \end{gather}
 where $J_n$ denotes the Bessel function of the first kind of the $n$th order, $y_{nm}$ is the $m$th positive root of the equation $J'_n(y)=0$, and $x_{nm}$ is the $m$th root of the equation $J_n(x) = 0$.
 \item Eigenvalues
 \be
 \omega_{nmk,\lambda(t)}^{\mathrm{TE}}=\sqrt{(\frac{y_{nm}}{R})^2+(\frac{\pi k_z}{\lambda(t)})^2},k_z\geq 1,
 \ee
 \be
 \omega_{nmk,\lambda(t)}^{\mathrm{TM}}=\sqrt{(\frac{x_{nm}}{R})^2+(\frac{\pi k_z}{\lambda(t)})^2},k_z\geq 0,
 \ee
 \item Coupled strengths
 \begin{gather}
  \begin{split}
  \label{a6}
 g^{\mathrm{TE}}_{nmk,n'm'k'}=\begin{cases}(-1)^{k+k'}\frac{2kk'}{k^2-k'^2}\delta_{mm'}\delta_{nn'}, &\text{if}\ k\neq k'\cr 0, &\text{if}\ k=k' \end{cases},
  \end{split}
 \end{gather}
 \begin{gather}
  \begin{split}
  \label{a7}
 g^{\mathrm{TM}}_{nmk,n'm'k'}=\begin{cases}(-1)^{k+k'}\frac{2k^2}{k^2-k'^2}\delta_{mm'}\delta_{nn'}, &\text{if}\ k\neq k'\cr \delta_{mm'}\delta_{nn'}, &\text{if}\ k=k' \end{cases}.
  \end{split}
 \end{gather}
 \end{enumerate}

 Similarly if the boundaries are at the radial coordinate $\rho=\lambda(t)$ and the longitudinal coordinate $z=(0,L_z)$,
 we have
 \begin{enumerate}
 \item Boundary conditions
 \be
 \left.\psi^{\mathrm{TE}}\right|_{z=0,L_z}=0,\ \ \ \ \left.(\partial_{\rho}+\dot{\lambda}(t)\partial_t)\psi^{\mathrm{TE}}\right|_{\rho=\lambda(t)}=0,
 \ee

 \begin{gather}
  \begin{split}
  \label{e21}
 \left.\partial_{z}\psi^{\mathrm{TM}}\right|_{z=0,L_z}=0,\ \left.\psi^{\mathrm{TM}}\right|_{\rho=\lambda(t)}=0,
  \end{split}
 \end{gather}

 \item Instantaneous orthonormal bases
 \begin{gather}
  \begin{split}
 \psi^{\mathrm{TE}}_{nmk,\lambda(t)}(\mathbf{r})=&\sqrt{\frac{2}{L_z}}\mathrm{sin}(\frac{\pi kz}{L_z})\\
                                        &\times\frac{\sqrt{2}J_n(y_{nm}\rho/\lambda(t))}{\lambda(t)J_{n}(y_{nm})\sqrt{1-n^2/y^2_{nm}}}\frac{e^{in\phi}}{\sqrt{2\pi}},
  \end{split}
 \end{gather}
 \begin{gather}
  \begin{split}
 \psi^{\mathrm{TM}}_{nmk,\lambda(t)}(\mathbf{r})=&\sqrt{\frac{2}{L_z}}\mathrm{cos}(\frac{\pi kz}{L_z})\ \ \ \ \ \ \ \ \ \ \ \ \ \ \ \ \ \ \ \ \ \ \ \ \ \ \ \ \ \ \ \ \ \ \ \\
                                        &\times\frac{\sqrt{2}J_n(x_{nm}\rho/\lambda(t))}{\lambda(t)J_{n+1}(x_{nm})}\frac{e^{in\phi}}{\sqrt{2\pi}},
  \end{split}
 \end{gather}
 \item Eigenvalues
 \be
 \omega_{nmk,\lambda(t)}^{\mathrm{TE}}=\sqrt{(\frac{y_{nm}}{\lambda(t)})^2+(\frac{\pi k_z}{L_z})^2},k_z\geq 1,
 \ee
 \be
 \omega_{nmk,\lambda(t)}^{\mathrm{TM}}=\sqrt{(\frac{x_{nm}}{\lambda(t)})^2+(\frac{\pi k_z}{L_z})^2},k_z\geq 0,
 \ee
 \item Coupled strengths
 \begin{gather}
  \begin{split}
  \label{a13}
 g^{\mathrm{TE}}_{nmk,n'm'k'}=\begin{cases}\frac{2y_{nm}y_{nm'}}{y^2_{nm}-y^2_{nm'}}\sqrt{\frac{y^2_{nm}-n^2}{y^2_{nm'}-n^2}}\delta_{nn'}\delta_{kk'}, &\text{if}\ m\neq m'\cr \frac{y^2_{nm}}{y^2_{nm}-n^2}\delta_{nn'}\delta_{kk'}, &\text{if}\ m=m' \end{cases},
  \end{split}
 \end{gather}
 \begin{gather}
  \begin{split}
  \label{a14}
 g^{\mathrm{TM}}_{nmk,n'm'k'}=\begin{cases}\frac{2x_{nm}x_{nm'}}{x^2_{nm}-x^2_{nm'}}\delta_{nn'}\delta_{kk'}, &\text{if}\ m\neq m'\cr 0, &\text{if}\ m=m' \end{cases}.\ \ \ \ \ \ \ \ \ \ \ \
  \end{split}
 \end{gather}
 \end{enumerate}

 In a spherical cavity, we introduce the spherical coordinates $(r, \theta, \phi)$. If the boundary is at the radial coordinate $r=\lambda(t)$,
 we have
 \begin{enumerate}
 \item Hertz potentials
 \begin{gather}
  \begin{split}
 \mathbf{\Pi}_e=\psi^{\mathrm{TM}}\hat{\mathbf{r}},
 \ \ \ \ \ \mathbf{\Pi}_m=\psi^{\mathrm{TE}}\hat{\mathbf{r}},
  \end{split}
 \end{gather}
 where we have used the Debye potentials~\cite{ph2006}.
 \item Boundary conditions
 \be
 \left.\psi^{\mathrm{TE}}\right|_{r=\lambda(t)}=0,
 \ee

 \begin{gather}
  \begin{split}
  \label{e21}
 \left.(\partial_{r}+\dot{\lambda}(t)\partial_t)(r\psi^{\mathrm{TM}})\right|_{r=\lambda(t)}=0,
  \end{split}
 \end{gather}
 \item Instantaneous orthonormal bases
 \begin{gather}
  \begin{split}
 \psi^{\mathrm{TE}}_{nlm,\lambda(t)}(\mathbf{r})=\sqrt{\frac{2}{\lambda^3(t)}}\frac{j_l(j_{ln}r/\lambda(t))}{j'_l(j_{ln})}Y_{lm}(\theta,\phi),\ \ \ \
  \end{split}
 \end{gather}
 \begin{gather}
  \begin{split}
 \psi^{\mathrm{TM}}_{nlm,\lambda(t)}(\mathbf{r})=\sqrt{\frac{2}{\lambda^3(t)}}\frac{j_l(\kappa_{ln}r/\lambda(t))}{j'_l(\kappa_{ln})\sqrt{\kappa^2_{ln}-l(l+1)}}Y_{lm}(\theta,\phi),
  \end{split}
 \end{gather}
  where $j_l$ denotes the spherical Bessel function of the $l$th order, $Y_{lm}(\theta,\phi)$ denotes the normalized spherical harmonics of degree $l$ and order $m$, $j_{ln}$ denotes the $n$th zero for $j_{l}(x)=0$, and $\kappa_{ln}$ denotes the $n$th zero of $\partial_x[xj_{l}(x)]=0$.
\item Eigenvalues
 \be
 \omega_{nlm,\lambda(t)}^{\mathrm{TE}}=j_{ln}/\lambda(t),
 \ee
 \be
 \omega_{nlm,\lambda(t)}^{\mathrm{TM}}=\kappa_{ln}/\lambda(t),
 \ee
 \item Coupled strengths
 \begin{gather}
  \begin{split}
  \label{a20}
 g^{\mathrm{TE}}_{nlm,n'l'm'}=\begin{cases}\frac{2j_{ln}j_{ln'}}{j^2_{ln}-j^2_{ln'}}\delta_{ll'}\delta_{mm'}, &\text{if}\ n\neq n'\cr 0, &\text{if}\ n=n' \end{cases},\ \ \ \ \ \ \ \ \ \ \ \ \ \ \ \ \ \ \
  \end{split}
 \end{gather}
 \begin{gather}
  \begin{split}
  \label{a21}
 g^{\mathrm{TM}}_{nlm,n'l'm'}=\begin{cases}\frac{2\kappa_{ln}\kappa_{ln'}}{\kappa^2_{ln}-\kappa^2_{ln'}}\sqrt{\frac{\kappa^2_{ln}-l(l+1)}{\kappa^2_{ln'}-l(l+1)}}\delta_{ll'}\delta_{mm'}, &\text{if}\ n\neq n'\cr \frac{\kappa^2_{ln}}{\kappa^2_{ln}-l(l+1)}\delta_{ll'}\delta_{mm'}, &\text{if}\ n=n' \end{cases}.
  \end{split}
 \end{gather}
 \end{enumerate}

 \renewcommand{\theequation}{B.\arabic{equation}}

 \setcounter{equation}{0}

 \section*{Appendix B: TM field in a trembling cavity}
 \label{Aa}

 The TM field in the trembling cavity satisfies different boundary conditions from the TE field. According to Refs.~\cite{qu2002,he2005}, the boundary conditions are
 \begin{gather}
  \begin{split}
  \label{ea1}
 &\left.\partial_{z}\psi^{\mathrm{TM}}\right|_{z=0}=0,\ \left.(\partial_{z}+\dot{\lambda}(t)\partial_t)\psi^{\mathrm{TM}}\right|_{z=\lambda(t)}=0,\\
 &\left.\psi^{\mathrm{TM}}\right|_{x=0,L_x}=\left.\psi^{\mathrm{TM}}\right|_{y=0,L_y}=0.
  \end{split}
 \end{gather}
 Similar to the TE field, we also define the following instantaneous orthonormal basis $\{\psi^{TM}_{\mathbf k,\lambda}({\mathbf r})\}$
 \begin{gather}
  \begin{split}
 \psi^{\mathrm{TM}}_{\mathbf{k},\lambda}(\mathbf{r})=&\sqrt{\frac{2}{\lambda(t)}}\mathrm{cos}(\frac{\pi k_z z}{\lambda(t)})\\
                                               &\times\frac{2}{\sqrt{L_xL_y}}\mathrm{sin}(\frac{\pi k_x x}{L_x})\mathrm{sin}(\frac{\pi k_y y}{L_y}),
  \end{split}
 \end{gather}
 which satisfies the Helmholtz equation (Eq.~\ref{eq:helmholtz}) and the boundary conditions (Eq.~(\ref{ea1})) to the first order of $\epsilon$ (notice $\left.\dot{\lambda}(t)\partial_t\psi^{\mathrm{TM}}_{\mathbf{k}}\right|_{z=\lambda(t)}\sim\epsilon^2$).
 Following the same procedure, we obtain the expression of the effective Hamiltonian, which is the same as that of the TE field (Eq.~(\ref{eq:H_eff})) except
 \begin{gather}
  \begin{split}
 g_{\mathbf{k}\mathbf{p}}=\begin{cases}(-1)^{k_z+p_z}\frac{2k_z^2}{k_z^2-p_z^2}\delta_{k_x p_x}\delta_{k_y p_y}, &\text{if}\ k_z\neq p_z\cr \delta_{k_x p_x}\delta_{k_y p_y}, &\text{if}\ k_z=p_z \end{cases}.
  \end{split}
 \end{gather}

 \renewcommand{\theequation}{C.\arabic{equation}}

 \setcounter{equation}{0}

\section*{Appendix C: the matrix representation technique}
\label{Ac}

 The matrix representation technique is a mathematical technique that can be used to calculate the trace of the products of several exponentials of a quadratic form in Boson operators. For simplicity, we first consider the product of two exponentials, i.e. $\hat J_1\hat J_2$, where $\hat J_i,i=1,2$ is an exponential of a quadratic form in Boson operator
 \be
 \label{a29}
 \hat{J}_i=\mathrm{exp}(\frac{1}{2}\hat{\bm{\alpha}}S_i\hat{\bm{\alpha}}).
 \ee
 Here, $\hat{\bm{\alpha}}=(\hat{a}_1,\hat{a}_2,\ldots,\hat a_n,\hat{a}^{\dag}_1,\hat{a}^{\dag}_2,\ldots,\hat a^\dag_n)$ and $\hat a_j,j=1,\ldots,n$ satisfies the bosonic commutation relations and the ($2n\times 2n$) matrix $S_i$ is a complex symmetric matrix. For later convenience, we introduce a characteristic matrix $[\hat{J}_i]$ corresponding to $\hat{J}_i$.
 \be
 \label{a30}
 [\hat{J}_i]=\mathrm{exp}(\sigma S_i),
 \ee
 where
 \be
 \sigma=\left(
 \begin{matrix}
 0&I\\
 -I&0
 \end{matrix}
 \right),
 \ee
 and $I$ is the $n\times n$ identity matrix. Now, let us define a set $\mathcal{\hat J}$ which includes all operators of the type (Eq.~(\ref{a29})) and the arbitrary product of these operators. Also, we define a set $\mathcal{J}$ which includes all matrices of the type (Eq.~(\ref{a30})) and the arbitrary product of these matrices. Then, it can be proved that both sets are the representations of the $2n$-dimensional complex symplectic group with operator and matrix multiplications respectively~\cite{no1969}. Also, the product of operators is related to the matrix multiplication, i.e. if $\hat{J}_3=\hat{J}_1\hat{J}_2$, we have $[\hat{J}_3]=[\hat{J}_1][\hat{J}_2]$, where $\hat J_i\in \mathcal{\hat J}$, and $[J_i]\in\mathcal{J},i=1,2,3.$ This result can be straightforwardly generalized to the product of more than two exponentials.
 So, the product of several exponentials of a quadratic form in Boson operators can be related to one exponential of a quadratic form in Boson operator of which the characteristic matrix is known.
 Finally, the trace of an exponential of a quadratic form in Boson operator can be calculated by its characteristic matrix~\cite{en1978} by
 \be
 \label{e29}
 \mathrm{Tr}\hat{J}=[(-1)^n \mathrm{det}([\hat{J}]-\tilde{I})]^{-1/2}.
 \ee
 where
\be
 \tilde{I}=\left(
 \begin{matrix}
 I&0\\
 0&I
 \end{matrix}
 \right).
 \ee
 Thus, we transform the trace of the product of several exponentials of a quadratic form in Boson operators to the calculation of the characteristic matrices which is easy to deal with.

 \renewcommand{\theequation}{D.\arabic{equation}}

 \setcounter{equation}{0}

\section*{Appendix D: analytical solutions for the general case with $\lambda_0 \neq \lambda_{\tau}$}
\label{Ad}

 The matrix representation technique can also be applied to more general cases in which the trembling boundary starts and stops at different positions $\lambda_0 \neq \lambda_{\tau}$. In the single-resonance conditions, the modified characteristic function ${\overline{\mathcal G}}(u,v)={\mathcal G}(u,v)e^{-iu\Delta \Phi}$ reads

 \begin{widetext}
 \begin{enumerate}
 \item The DoF resonance: $\Omega = 2\omega_{\mathbf k}$
 \begin{eqnarray}
 {\overline{\mathcal G}}_{1}(u,v)=\frac{\mathrm{sinh}\frac{-iu\Delta\omega_{\bm{k}}+\beta\omega_{\bm{k},\lambda_0}}{2}}{\sqrt{\mathrm{sinh}^2\frac{-iu\Delta\omega_{\bm{k}}+\beta\omega_{\bm{k},\lambda_0}}{2}+\mathrm{sin}(u\omega_{\bm{k},\lambda_\tau}+v)\mathrm{sin}((u-i\beta)\omega_{\bm{k},\lambda_0}+v)\mathrm{sinh}^2g_1\tau}},\\
 \end{eqnarray}
\item The SuF resonance: $\Omega = \omega_{\mathbf k}+\omega_{\mathbf p}$
\begin{eqnarray}
 {\overline{\mathcal G}}_{2}(u,v)=\frac{\mathrm{sinh}\frac{-iu\Delta\omega_{\bm{k}}+\beta\omega_{\bm{k},\lambda_0}}{2}\mathrm{sinh}\frac{-iu\Delta\omega_{\bm{p}}+\beta\omega_{\bm{p},\lambda_0}}{2}}{\mathrm{sinh}\frac{-iu\Delta\omega_{\bm{k}}+\beta\omega_{\bm{k},\lambda_0}}{2}\mathrm{sinh}\frac{-iu\Delta\omega_{\bm{p}}+\beta\omega_{\bm{p},\lambda_0}}{2}+\mathrm{sin}\left(\frac{u(\omega_{\bm{k},\lambda_\tau}+\omega_{\bm{k},\lambda_\tau})}{2}+v\right)\mathrm{sin}\left(\frac{(u-i\beta)(\omega_{\bm{k},\lambda_0}+\omega_{\bm{p},\lambda_0})}{2}+v\right)\mathrm{sinh}^2g_2\tau},
 \end{eqnarray}
\item The DiF resonance: $\Omega = \left|\omega_{\mathbf k}-\omega_{\mathbf p}\right|$
\begin{eqnarray}
 {\overline{\mathcal G}}_{3}(u,v)=\frac{\mathrm{sinh}\frac{-iu\Delta\omega_{\bm{k}}+\beta\omega_{\bm{k},\lambda_0}}{2}\mathrm{sinh}\frac{-iu\Delta\omega_{\bm{p}}+\beta\omega_{\bm{p},\lambda_0}}{2}}{\mathrm{sinh}\frac{-iu\Delta\omega_{\bm{k}}+\beta\omega_{\bm{k},\lambda_0}}{2}\mathrm{sinh}\frac{-iu\Delta\omega_{\bm{p}}+\beta\omega_{\bm{p},\lambda_0}}{2}+\mathrm{sin}\frac{u(\omega_{\bm{k},\lambda_\tau}-\omega_{\bm{k},\lambda_\tau})}{2}\mathrm{sin}\frac{(u-i\beta)(\omega_{\bm{k},\lambda_0}-\omega_{\bm{p},\lambda_0})}{2}\mathrm{sin}^2g_3\tau},
 \end{eqnarray}
 \end{enumerate}
 \end{widetext}
 where $\Delta\omega_{\bm{s}}=\omega_{\bm{s},\lambda_\tau}-\omega_{\bm{s},\lambda_0},\bm{s}={\mathbf k},{\mathbf p}.$

 It is worth mentioning that the Crook's fluctuation theorem in these cases is non-trivial. For later convenience, we consider the moving boundary takes a more general form
 \be
 \lambda(t)=\lambda_0[1+\epsilon\mathrm{sin}(\Omega t+\varphi)],
 \ee
 where $\varphi$ denotes the initial phase of the boundary. In this case, the effective Hamiltonian after the RWA becomes $e^{-\frac{i\varphi}{2}\hat N_{\lambda_{0}}}\hat V^{\mathrm I}e^{\frac{i\varphi}{2}\hat N_{\lambda_{0}}}$. The characteristic function will not change after the unitary transformation because of the cyclic property of trace.

 In the nonzero initial phase $\varphi$ cases, the modified characteristic functions associated with the forward process $\lambda(t)$ and the reverse process $\lambda(\tau-t)$ read
 \be
 {\overline{\mathcal G}}_{F,j}(u,v)={\overline{\mathcal G}}_{j}(u,v),j=1,2,3.
 \ee

 \begin{gather}
  \begin{split}
 {\overline{\mathcal G}}_{R,1}(u,v)={\overline{\mathcal G}}_{1}(u,v)|_{\omega_{\bm{k},\lambda_0}\leftrightarrow\omega_{\bm{k},\lambda_\tau}},\\
 {\overline{\mathcal G}}_{R,j}(u,v)={\overline{\mathcal G}}_{j}(u,v)|_{\omega_{\bm{k},\lambda_0}\leftrightarrow\omega_{\bm{k},\lambda_\tau}\atop\omega_{\bm{p},\lambda_0}\leftrightarrow\omega_{\bm{p},\lambda_\tau}},j=2,3,
  \end{split}
 \end{gather}
 where $f|_{a\leftrightarrow b}$ means the value of the parameters $a$ and $b$ in function $f$ is interchanged. Then, we have ${\mathcal G}_{\rm R}(-u,-v)={\mathcal G}_{\rm F}(u+i\beta,v-i\beta\mu)e^{\beta\Delta\Phi}$,
 which is equivalent to the Crook's fluctuation theorem $\frac{{\mathcal P}_{\rm F}\left(w,\Delta N\right)}{{\mathcal P}_{\rm R}\left(-w,-\Delta N\right)}=e^{\beta\left(w-\nu\Delta N-\Delta\Phi\right)}$ and the time reversal symmetry of the effective Hamiltonian $\hat H_{\mathrm{eff}}(t)$~\cite{th2009}
 \be
 \mathcal{\hat{T}}^{-1}\hat{H}_{\mathrm{eff}}(\lambda(t),\dot{\lambda}(t))\mathcal{\hat{T}}=\hat{H}_{\mathrm{eff}}(\lambda(t),-\dot{\lambda}(t)),
 \ee
 where the time-reversal operator $\mathcal{\hat{T}}$ is an anti-linear and anti-unitary operator and which implies changing the sign
 of all odd operators. We would like to emphasize that $\mathcal{\hat{T}}^{-1}\hat{\psi}^{\rm TE}(\bm r, t)\mathcal{\hat{T}}=-\hat{\psi}^{\rm TE}(\bm r, -t)$ and $\mathcal{\hat{T}}^{-1}\hat{\psi}^{\rm TM}(\bm r, t)\mathcal{\hat{T}}=\hat{\psi}^{\rm TM}(\bm r, -t)$, which indicate that the electric field $\bm E$ (the magnetic field $\bm B$) is an even (odd) operator.
 It is interesting to notice that the time derivative of the work parameter $\dot{\lambda}(t)$ also appears in the effective Hamiltonian in Ref.~\cite{re2017}.


\begin{thebibliography}{99}
\bibitem{footnote1} Only the processes of even photons are allowed in our system, which is a manifestation of the $\mathbb Z_2$ symmetry, $\hat{\psi}(x,t)\rightarrow-\hat{\psi}(x,t)$, and the conserved quantity, $\mathrm{exp}(i \pi \sum_{\bm{k}}\hat{a}_{\bm{k}}^\dag\hat{a}_{\bm{k}})$.
\bibitem{footnote3} Please note that the limit $\hbar\rightarrow0$ is equivalent to the limit $\beta\rightarrow0$ meanwhile keeping $u/\beta$ as a constant.
\bibitem{footnote5} If we are interested in the marginal distribution of work only, we can set $v=0$ (see Eq.~(\ref{eq:marginal_PW})).
\bibitem{footnote2} This conclusion can be considered as the quantum-classical correspondance between a photon field and a classical EM field. Because there are infinite number of modes in the black-body radiation, the thermal equilibium state of a classical EM field is not well-defined. However, in our current study because only a few resonant modes are involved, we can define the thermal equilibium state in the subspace spanned by these modes. And the quantum-classical correspondance is straightfoward since every mode of the EM field is related to a harmonic oscillator.
\end{thebibliography}
  \end{document}